\documentclass[lettersize,journal]{IEEEtran}
\usepackage{amsmath,amsfonts}
\usepackage{algorithmic}
\usepackage{algorithm}
\usepackage{array}
\usepackage[caption=false,font=normalsize,labelfont=sf,textfont=sf]{subfig}
\usepackage{textcomp}
\usepackage{stfloats}
\usepackage{url}
\usepackage{verbatim}
\usepackage{graphicx}
\usepackage{cite}
\usepackage{hyperref}
\usepackage[table]{xcolor}
 \usepackage{makecell, cellspace, caption}
\usepackage{array,multirow,tabularx,ragged2e, booktabs}
\usepackage{pifont}
\usepackage{fontawesome}
\newcommand{\cmark}{\ding{51}}
\newcommand{\xmark}{\ding{53}}
\usepackage{multirow}

\begin{document}

\title{Towards Trusted and Intelligent Cyber-Physical Systems: A Security-by-Design Approach}
\author{Sabah Suhail and Raja Jurdak,~\IEEEmembership{Senior~Member,~IEEE} 
\thanks{S. Suhail is with Vienna University of Economics and Business, Vienna, Austria (e-mail: sabah.suhail@wu.ac.at).}
\thanks{R. Jurdak is with Queensland University of Technology, Australia (e-mail: r.jurdak@qut.edu.au).}
}



\maketitle

\begin{abstract}
The complexity of cyberattacks in Cyber-Physical Systems (CPSs) calls for a mechanism that can evaluate the operational behaviour and security without negatively affecting the operation of live systems.
In this regard, Digital Twins (DTs) are revolutionizing the CPSs. DTs strengthen the security of CPSs throughout the product lifecycle, while assuming that the DT data is trusted, providing agility to predict and respond to real-time changes.
However, existing DTs solutions in CPS are constrained with untrustworthy data dissemination among multiple stakeholders and timely course correction. Such limitations reinforce the significance of designing trustworthy distributed solutions with the ability to create actionable insights in real-time. To do so, we propose a framework that focuses on trusted and intelligent DT by integrating blockchain and Artificial Intelligence (AI). Following a hybrid approach, the proposed framework not only acquires process knowledge from the specifications of the CPS, but also relies on AI to learn security threats 
based on sensor data. Furthermore, we integrate blockchain to safeguard product lifecycle data. We discuss the applicability of the proposed framework for the automotive industry as a CPS use case. Finally, we identify the open challenges that impede the implementation of intelligence-driven architectures in CPSs.
\end{abstract}


\section{Introduction}\label{introduction}
\IEEEPARstart{I}{n} the paradigm of Industry 4.0, traditional Industrial Control Systems (ICSs) that include Operational Technology (OT) are increasingly integrated with general-purpose Information Technology (IT) systems~\cite{dietz2020unleashing}.
Whereas integrating IT and OT systems in Cyber-Physical Systems (CPSs) provides promising solutions in a plethora of ubiquitous industrial ecosystems, it introduces novel attack vectors~\cite{suhail2021blockchainbased} such as Man-in-the-Middle (MITM). Usually, in IT/OT-system operational functionality outweighs security~\cite{dietz2020unleashing}. Furthermore, if addressed at all, security is often added retrospectively during the operation phase of the asset rather than at the initial design phase. 
Thus, loopholes in the system infrastructure enable attackers to gain deep knowledge of system behavior and launch covert attacks or Advanced Persistent Threats (APTs). Considering human-machine collaboration scenarios, such attacks not only degrade the overall system performance but may pose human safety risks. One such example of ICS-tailored malware is Stuxnet~\cite{langner2013kill}, where, without being detected, a malicious code can intercept and modify the data sent to and from Programmable Logic Controllers (PLCs).


Considering specifically designed covert and long-term attacks like APTs that access and exfiltrate information from systems, essential measures to fortify CPS involve: (1) evaluating the functional behaviour of the system, and (2) conducting security attacks on the system to identify vulnerabilities or threats. Since CPSs can not be deactivated for carrying out such analysis, assessing the system's security level calls for online solutions that accurately reflect the actual CPS in operation while avoiding any interference or implications of testing on the live systems. In this regard, Intrusion Detection System (IDS) (signature- or anomaly-based) is a promising approach to uncover malicious activities by identifying deviations or patterns from a defined benign behavior~\cite{LIAO201316}. However, false alarms may raise human safety concerns, and attack scenarios are hard to reproduce~\cite{eckhart2018towards}. Another potential solution involves deploying testbeds as a controllable cyber prototype for testing the functional requirements and operational behavior of the system; nevertheless, a testbed setup and maintenance can be time-consuming and cost-intensive~\cite{Eckhart2019}. Most importantly, attackers may harm the CPS infrastructure as mere detection of attacks by testbed-based solutions often leads to delayed countermeasures. Moreover, such solutions may not cover the entire lifecycle of the product (including design and development, operation and maintenance, and decommissioning).

\begin{figure}[!ht]
\centerline{\includegraphics[width=3.0in]{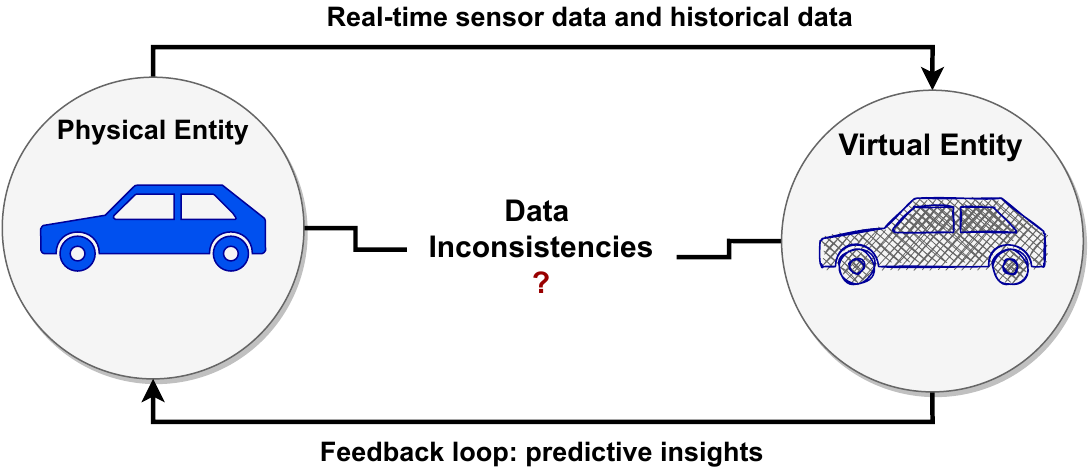}}
\caption{Building digital twins: A simple illustration.}
\label{fig:DTmodel}
\end{figure}

To address the above-mentioned issues requires a solution with the following characteristics: (i) able to analyze and optimize the physical processes within CPS before their real-world implementation, (ii) can evaluate and connect the entire product lifecycle, and (iii) can prevent and detect security loopholes in the CPS without obstructing the ongoing operations.
Digital Twins (DTs) are considered one such solution.
Driven by asset-centric data, DTs are virtual replicas of physical systems that mirror every facet of a product or process and can provide actionable insights through monitoring, optimization, and prediction~\cite{tao2019make, tao2018digitalsurvey}. 
During industrial physical processes, DTs collect and integrate data from multiple sources, such as sensor and actuator data from the factory floor, historical production data derived from product lifecycle data, and domain knowledge to generate outputs in the form of models, simulations, replications, or behavioral analytics. Following a closed-loop, the DTs inspect for data inconsistencies between the physical entity and virtual entity (as shown in Fig.~\ref{fig:DTmodel}), and feedback the simulation data to the physical entity to adopt better calibration and testing strategies. Such recurrent processes evolve DT models and their physical counterparts to support more accurate estimation, prediction, and optimization of the industrial processes. 

The lifecycle of DTs spans different phases where multiple stakeholders (i.e., owners, manufacturers, distributors, and maintainers) carry out various operations~\cite{suhail2021blockchainbased}. Some of these critical operations require recording changes to detect deviations throughout the process cycle, for instance, trustworthy state of twin software~\cite{gehrmann2019digital}, update commands to PLCs~\cite{hadvziosmanovic2014through}, or time-dependent modification to the process variables~\cite{patel2021real}. The disparate data silos act as a barrier in orchestrating the data in a digital thread that requires linking DT data throughout various phases of the product lifecycle~\cite{suhail2021blockchainbased}, therefore, the centralized data management architectures are not well suited for such environments. Additionally, it is essential to capture the trustworthiness of data as the accumulated data accounts for rational decision making. 
Enabling the distributed storage of information with no single point-of-control can be established through blockchain, which can facilitate data provenance. Moreover, provenance-enabled blockchain-based digital twins can aid in identifying the faulty node in the network. Thus, combining DTs and blockchain can reshape the CPSs where blockchain acts as a trustworthy data dissemination source to physical processes while DTs extract actionable insights for predictive maintenance and situational-awareness~\cite{suhail2020trustworthy}.

\begin{table*}[!ht]
\centering
\caption{Information security domain: Obtaining process knowledge for digital twins.} \label{tab:related_works}
\renewcommand*{\arraystretch}{1.5}
\tabcolsep=0.2cm
\begin{tabular}{|p{2.5cm} | p{1.0cm}| p{1.0cm}| p{1.0cm} |p{1.0cm}| p{1.0cm}| p{1.0cm}| p{1.5cm}| p{4.0cm}|}
\hline
{\thead{\textbf{Process}\\ \textbf{knowledge}}} & \textbf{Ref.} & \textbf{Data variety} & \textbf{Trusted sources} & \textbf{Data fusion} &  \textbf{S\&S} &  \textbf{Data storage} &  {\thead{\textbf{Framework}/\\\textbf{POC}/\\\textbf{Emulation}}} & \textbf{Objective} \\
\hline
\hline

\multirow{6}{*}{Specification-based}
& \cite{eckhart2018towards}  & \xmark  & \faAdjust  & \xmark  & \cmark & \xmark & POC & Detect rule violations \\
& \cite{eckhart2018specification}  & \faAdjust  & \faAdjust  & \faAdjust & \cmark & \faAdjust & POC & State-based intrusion detection \\
& \cite{eckhart2019enhancing}  & \xmark  & \faAdjust  & \xmark & \cmark & \xmark & Framework & Risk assessment, Incident handling  \\
& \cite{dietz2020integrating}  & \xmark  & \xmark  & \xmark & \cmark & \xmark & POC & Incident analysis\\
& \cite{dietz2019distributed}   & \cmark  & \xmark  & \faAdjust & \faAdjust & \cmark & Framework & Secure data sharing\\
& \cite{patel2021real}   & \cmark  & \xmark  & \xmark & \faAdjust & \cmark & Emulation & Process-aware intrusion detection \\

\hline
\multirow{2}{*}{ML-based} 
& \cite{groshev2021toward} & \xmark & \xmark & \xmark  & \xmark & \faAdjust & POC & Intrusion
detection \\
&\cite{damjanovic2018digital} & \xmark & \xmark  & \xmark & \xmark & \xmark  & Framework & Vulnerability detection \\
\hline
Hybrid & Proposed & \cmark & \cmark & \cmark & \cmark & \cmark & Framework &  Situational awareness\\
\hline
\end{tabular}%

\xmark~Not covered; \faAdjust~Partially covered; \cmark~Covered; Proof of Concept (POC)
\end{table*}

For greater agility, industrial processes demand automation to trigger timely and seamless actions. 
Considering the dynamism of CPSs, the overarching question is \emph{how to automate cyber-physical situational awareness while providing online threat intelligence (TI)?} 
TI provides information on attacker behaviour that security analysts can use to recognize the indicators of compromise (IOCs) on time~\cite{bouwman2020different}. To ensure sustainable protection against attacks, it is necessary to automate extracting TI and evidence-based insights from data sources. While keeping up with the complexity and adaptability of current cybersecurity threats, Artificial Intelligence (AI)-driven security solutions provide automated security solutions to detect zero-day threats~\cite{ibrahim2020challenges}. In this regard, we utilize intelligence-driven solutions (such as data analytics and TI) in DTs that can help to identify existing vulnerabilities, faulty models, malicious actors, and potential attack vectors, thereby minimizing the threat landscape and improving the CPSs.
In the information security domain, the concept of building the process knowledge of DTs can be seen from two perspectives. Firstly, utilizing CPS specification to model the physical counterpart based on engineering artifacts~\cite{eckhart2018towards, eckhart2019enhancing, eckhart2018specification}. Secondly, utilizing Machine Learning (ML) methods to learn security-related aspects based on sensor data~\cite{groshev2021toward, damjanovic2018digital} without obtaining process knowledge from DTs. 
Both solutions lack either real-time breach detection or potential ability to model the correct behavior of physical counterparts. Furthermore, present works do not consider whether data sources are trustworthy, which is critical to ensure the quality of input data to DTs. Similarly, the need for data storage and data integration and interoperability are also not given due attention while proposing the frameworks or POCs (as shown in Table~\ref{tab:related_works}). 

\begin{figure}[!t]
\centerline{\includegraphics[width=3.0in]{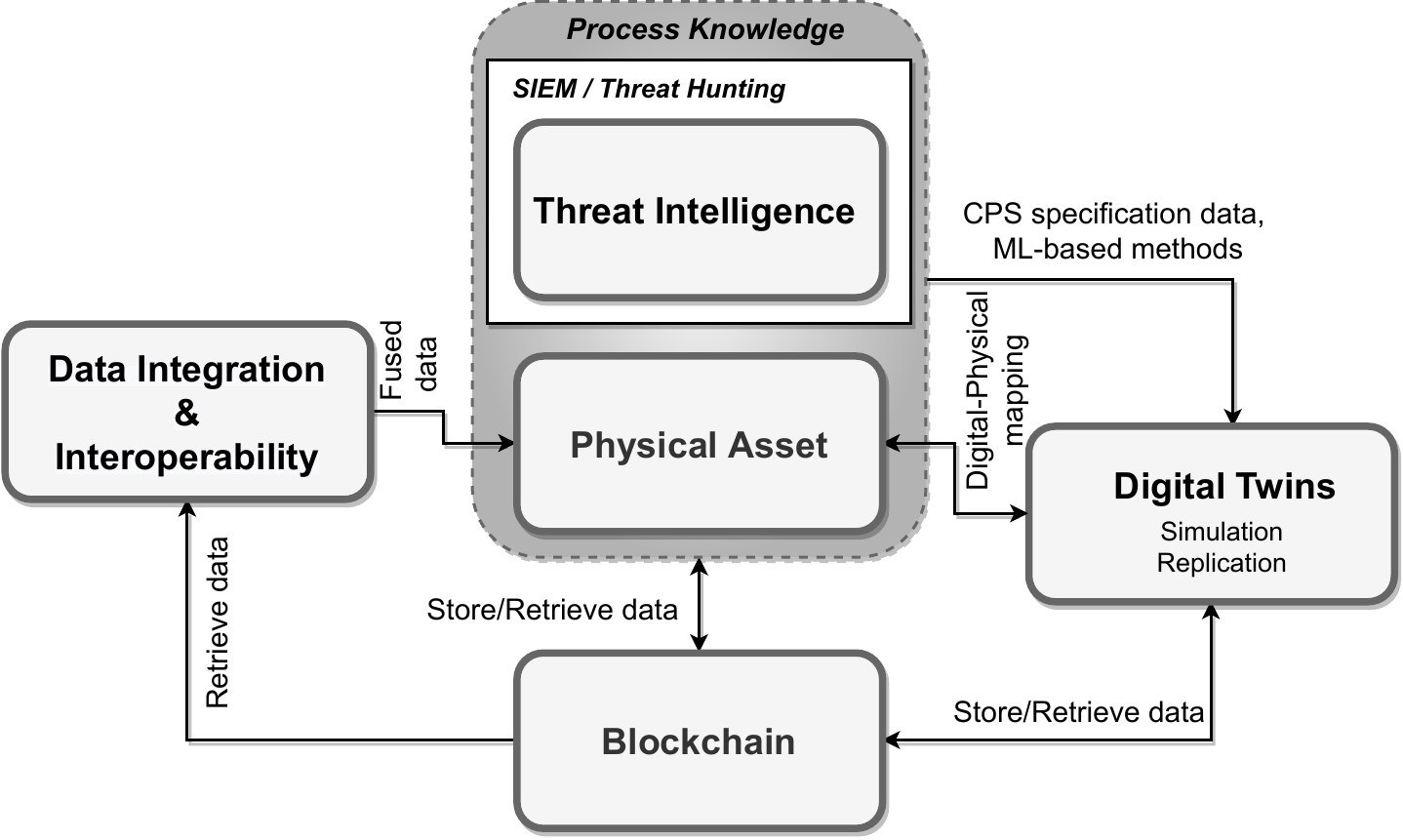}}
\caption{Overview of the proposed framework.}
\label{fig:abstract}
\end{figure}

To address the above limitations, this article proposes an AI-aided blockchain-based DT framework (shown in Fig.~\ref{fig:abstract}) to secure CPSs. Our framework adopts a hybrid approach to build DTs, where in addition to acquiring process knowledge from the specification of the CPS on the logic/network layer, we also rely on TI to learn and prevent attacks. Most importantly, we integrate blockchain to safeguard product lifecycle data.

 Our main contributions are as follows:



\begin{itemize}
\item To secure CPSs, we adopt a hybrid approach to build DTs, where we acquire process knowledge from the specification of the CPS and TI to learn and prevent attacks.
\item We leverage blockchain for safeguarding product lifecycle. Furthermore, recording/retrieving data from blockchain ensure trustworthy environment for CPS and DTs, which act as a source of input data for CPS.
\item We discuss the feasibility of the proposed framework for the automotive industry as a CPS use case.
\end{itemize}

The rest of the paper is organized as follows: Section~\ref{sec:related-works} covers the related work in the context of creating DTs for situational awareness. Motivated by the security challenges of the CPSs, we propose a framework in Section~\ref{sec:framework}. Section~\ref{sec:use_case} maps the proposed framework to an automotive industry use case. Finally, Section~\ref{conclusion} concludes the paper with future research directions.

\section{Related Work} \label{sec:related-works}
This section primarily focuses on the research that covers the initial development phase of DTs, which comprises of obtaining process knowledge either through system specification and/or ML-based approaches. Most importantly, we narrow down the existing works based on the objective of using DTs, i.e., an enabler of cyber situational awareness (state-based or process-based). Table~\ref{tab:related_works} summarizes the comparative analysis of the existing works.

The specification-based approaches (for instance,~\cite{eckhart2018towards, eckhart2018specification, dietz2019distributed, dietz2020integrating, patel2021real}) leverage the technical, topological, and control artifacts of the underlying system that are maintained throughout the system engineering process. 
With a traceable and well-documented system specification support, this strategy of building DTs provides an automatic and scalable solution for generating a virtual environment~\cite{eckhart2018towards}. 
However, in comparison to ML-based solutions, they are not able to create future scenarios. In the existing literature, most of the papers in this category focus on the initial development of DTs. However, the consideration of data variety, data trustworthiness, data storage/retrieval, and data fusion are missing.  
In case of ML-based approaches (for instance,~\cite{groshev2021toward, damjanovic2018digital}), DTs are not aware of any control logic per se as 
process knowledge can merely be learned through ML-based methods~\cite{Eckhart2019}.

The works on blockchain-based DTs~\cite{khan2020towards, suhail2021blockchainbased, mandolla2019building, putz58ethertwin} cover various aspects of traceability, storage, retrieval, and sharing of DT data among the involved parties in addition to other security aspects (such as quantum resistance, encryption, access control, etc.). Other than~\cite{dietz2019distributed}, the other blockchain-based schemes lack the details on how the DTs are constructed and how DT security operation modes can be used to create future scenarios in the absence of direct observation data. Furthermore, they vary in terms of utilizing DTs and are therefore beyond the scope of this paper. However, our previous work in~\cite{suhail2021blockchainbased} provides a comprehensive review of the design and implementation issues in blockchain-based DTs. 

Without considering the requirement of data storage, data integration, and interoperability, both types of solutions lack either real-time breach detection or the potential ability to exploit the correct behaviour of physical counterparts (as shown in Table~\ref{tab:related_works}). 
Considering both solutions, we adopt a hybrid approach to build DTs. In addition to acquiring process knowledge from the specification of the CPS, we also rely on TI to learn and prevent attacks. Moreover, to safeguard the product lifecycle data, a blockchain-based solution is integrated.

\section{Towards Trusted and Intelligent Cyber-Physical Systems} \label{sec:framework}
\begin{figure*}[!t]
\centerline{\includegraphics[width=4.5in]{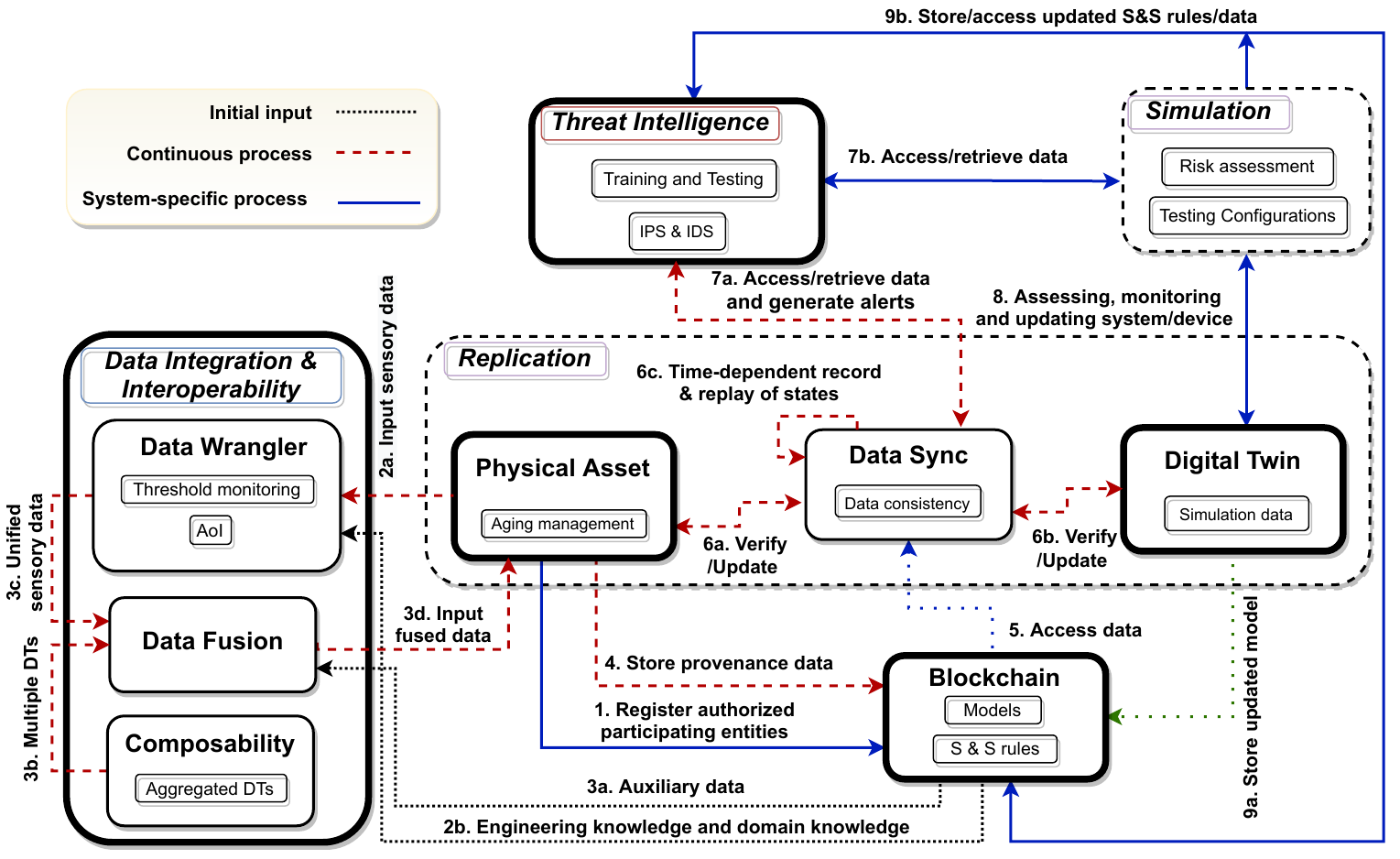}}
\caption{Securing Cyber-Physical Systems: A threat intelligence-aided blockchain-based digital twin framework.}
\label{fig:framework}
\end{figure*}

This section overviews the components of the framework for secure CPSs. The framework presented in Fig.~\ref{fig:framework} comprises four key components including (i) data integration and interoperability, (ii) physical asset and its replica, (iii) blockchain ledger, and (iv) threat intelligence (TI).
The {\it data integration and interoperability} is responsible for cleaning invalid, duplicate, or missing data, converting heterogeneous data formats into a unified data structure, and aggregating data to generate a consistent interpretation of a certain object before inputting data into the physical space.
The {\it physical asset} and its {\it clone} counterpart must be constantly connected to continuously perform the digital-physical mapping between the predefined system performance parameters retrieved from the storage, and real-time sensor data from the manufacturing unit, to verify data consistency. 
The {\it blockchain ledger} enforces secure data management by storing data and recalling events.
{\it TI} identifies vulnerabilities, threats, existing and potential attack vectors to minimize the threat landscape. The TI module can be further utilized by Security Information and Event Management (SIEM)~\cite{dietz2020integrating} and/or threat hunting.
The DTs are build by acquiring process knowledge from the specifications of CPS and TI.


Fig.~\ref{fig:framework} represents the detailed working of the main components presented in Fig.~\ref{fig:abstract}. We divide the data flow throughout the system into three types of processes: (1) the initial input is provided to the data wrangler and data fusion methods; (2) to stay consistent with the requirement of the underlying application, the continuous process updates the data or system state based on a defined time period; and (3) the system-specific process is scheduled based on the underlying events or triggered depending on cyber situational awareness of the CPSs. 
The steps shown in Fig.~\ref{fig:framework} can be summarized as follows: The participating entities (such as sensors, machines, and humans) register as authorized entities on the blockchain ({\it step 1}). The multimodal and heterogeneous sensors or actuator data from physical assets ({\it step 2a}) and system-specific input knowledge such as engineering and domain knowledge ({\it step 2b}) are cleaned and converted to a unified format at the data wrangler. The data from multiple sources, for instance, auxiliary data ({\it step 3a}), multiple DT instances data ({\it step 3b}), and unified sensory data ({\it step 3c}) is aggregated by data fusion and is inputted to the physical asset to support consistent and comprehensive representation of data ({\it step 3d}). To facilitate track and trace solutions, the provenance data is recorded on the ledger ({\it step 4}). Provenance records a complete lineage of data along with a set of actions performed on data~\cite{ZAFAR201750}. 
The DTs have two security operation modes, i.e., simulation and replication that support monitoring and replaying of CPS events. 
In the replication mode, the states or events from the physical environment are continuously recorded ({\it step 6a}) and are periodically replayed at DTs ({\it step 6b}). The purpose of this operation is to track data inconsistencies (as shown in Fig.~\ref{fig:DTmodel}). To do this, an time-dependent digital-physical mapping is performed using the data sync method {\it step 6c}. During the process, any essential data (such as Safety \& Security (S\&S) rules) can be accessed by the blockchain ({\it step 5}). S\&S rules define thresholds and consistency checks to detect attacks and thus can aid in tracking rule violations. In the simulation mode, the states or events are monitored independently of the physical space and can be used to update the twin ({\it step 8}) and eventually the physical asset. 
For preventing eventual failures that may be caused by security breaches or suspicious flows, the future state of a physical asset must be forecast to trigger necessary actions, such as maintenance. Therefore, to effectively use sensors real-time data streams in DTs, TI accesses data from data sync (through replication mode {\it step 7a}) and DT (through simulation mode ({\it step 7b}). The ML-based solutions can preemptively detect failures and call the corresponding scheduling services (in the physical space) or model calibration services (in the virtual space) to carry out the necessary measures such as configuring the machine settings based on tool wear data or tuning model parameters to simulate the physical counterpart with high fidelity.
The updated models upon calibration ({\it step 9a}) and S\&S rules must be stored and retrieved from the ledger to ensure their reliability ({\it step 9b}) and thus strengthen the rationale for integrating blockchain with DTs.

In the following, we discuss each component of the TI-aided blockchain-based DT framework in detail. 

\subsection{Data Integration and Interoperability} \label{DDI}
{\textbf{Challenge \#1:}} {\it How to avoid the insertion of erroneous data into the physical asset and its twin?}

The integration and interoperability of data from multimodal and heterogeneous sensors are significant prerequisites of DT implementation. 
This process begins by inputting sensory data from physical assets to the {\it data wrangler}. In our proposed framework, the data wrangler is responsible for cleaning invalid, duplicate, or missing data, converting heterogeneous data formats into a unified data structure. AI-enabled data curation helps in improving the data quality by implementing intelligent automated methods to fill in the missing data and to clean the data (such as data denoising, data de-duplication)~\cite{suhail2021blockchainbased}.
While stringent security guarantees are inherited from the blockchain, ensuring the~\emph{trustworthiness} of data-generating sources is equally important for critical infrastructures. Because the data is input to other entities (such as DTs, TI) to make decisions, ignoring such measures may lead to the insertion of false data into the system resulting in Garbage-In, Garbage-Out (GIGO) problem~\cite{suhail2020trustworthy}. Therefore, three-fold Integrity Checking Mechanisms (ICMs) ({\it engineering knowledge and domain knowledge}) is established by: (i) collecting data only from registered devices; (ii) cross-validating the device data against threshold values; and (iii) minimizing the Age of Information (AoI) based on time elapsed since the generation of latest status update
received at destination.
Given that blockchain mechanisms do not guarantee the trustworthiness of data at the origin, we can rely on utilizing a blockchain-based layered trust architecture~\cite{dedeoglu2019trust} for sensor data cross-validation.
Engineering knowledge describes the design specifications of the underlying CPS at the network/logic layer: (i) system-level components (specifying device configuration details and control logic), (ii) network-level information (specifying the topology and communication path through logical connections and endpoints), and (iii) the relationship among components (specifying process-level data aggregation, fine-grained policies and constraints)~\cite{eckhart2018towards}. 
Such explicit definitions of the technical, topological, and control artifacts can help to generate the network setup of the virtual environment~\cite{eckhart2018towards}. Furthermore, such engineering knowledge can also serve as a basis for implicit security rules. For instance, defining a safe state based on normal operations of devices or specific services provided by the process, deriving a whitelist from network-level monitoring based on the authorized addressing and routing information, detecting unknown devices or unidentified connections, identifying abnormal changes in the control logic, etc~\cite{eckhart2018towards}. In addition to these measures, Internet of Things (IoT) sensors are calibrated periodically for ageing management and fault diagnosis at the initial stages of data collection. 
In addition to specifying engineering knowledge, the framework also includes domain-specific knowledge from experts in various fields such as engineers (electrical, mechanical, instrumentation, and control), supply chain entities, security professionals, etc. Once generated, the domain knowledge can be used as a reference by different organizations and can be tailored to meet their specific needs.

Next, to facilitate an abstract view of the overall phenomena, DTs must support the correlation of different instances of DTs associated with different physical sub-components. 
For this purpose, multiple instances of DTs, where each instance mirrors a physical asset and/or activity, can be aggregated at {\it composability} to mimic the bigger picture of the physical world. Moreover, auxiliary data (user- or application-specific data) can also be provided. Finally, the fused data is inputted to the physical asset to carry out the physical-digital mapping through data sync. Thus, data interoperability and integration provide a more consistent, comprehensive, and accurate representation than the single perspective of data. 

\subsection{Digital Twins}
{\textbf{Challenge \#2:}} {\it How to accurately reflect the behavior of CPS while operating virtually in an isolated environment disjointed from live systems throughout the product lifecycle?}

DT is a virtual representation of a real-world system that utilizes real-time and historical data to analyze, predict, and optimize operations without interfering with the actual environment and with lifecycle security. DTs have two operation modes (i) replication and (ii) simulation. 
The replication mode provides digital tracing of real-world
events by mirroring data from the physical environment. For replication mode, DTs and their physical counterparts must be {\it constantly connected} in a sense that the virtual replica must continuously reflect the physical object through log files, sensor measurements, network communication, etc. Depending on the application requirements, data can be collected from the physical space after a certain time interval or even offline. 
To facilitate the {\it trial and error} approach, the simulation mode runs independently of its physical counterpart. In addition to being reproducible, this mode allows running tests repeatedly by resetting the model through a broad range of specified conditions to support the comprehension of emergent system behavior. The simulation mode supports {\it security by design} approach through which it facilitates analyzing process changes, test devices, or detect misconfigurations by performing security tests within the virtual environment.


\subsection{Blockchain Empowered Digital Twins}
{\textbf{Challenge \#3:}} {\it How to ensure integrity and trustworthiness of data collected from the physical world that are then fed into the DTs?} 

Although DTs provide powerful means to control, govern, and program the lifecycle of physical resources for supporting the provision of product servitization to end-users, these benefits of DTs are based on an assumption about data trust, integrity, and security. Data trustworthiness is considered to be more critical regarding the integration and interoperability of multiple components or subcomponents among different DTs to provide an aggregated view of the complex physical system. Nevertheless, in real-life scenarios, data breaches could occur due to several reasons, both maliciously or mistakenly~\cite{suhail2021blockchainbased}. Therefore, curating and mining actionable insights from the collected data calls for a data storage infrastructure that can manifest the dissemination of trustworthy and secure data~\cite{suhail2020trustworthy}. In this regard, provenance-enabled blockchain-based DTs can facilitate digital identity and data traceability from disparate data repositories while reasoning about the current state and the chained actions on a data object (such as~\textit{who},~\textit{when},~\textit{where}, and~\textit{how}), thereby ensuring trustworthy DTs throughout the product lifecycle~\cite{suhail2020trustworthy}.

Considering the infrastructure-related blockchain design challenges, there exists no ideal solution that efficiently supports CPSs ranging from throughput, scalability, storage efficiency, and security. To gain practical insights in CPSs, blockchain-related design challenges may cover deployment of blockchains, blockchain types, consensus mechanisms, scalability, and storage. 
The deployment of blockchain may vary based on the underlying architecture. Considering resource-constrained IoT devices, deploying blockchain on fog or edge nodes to carry out resource-intensive tasks (such as running a consensus algorithm or executing smart contracts) can provide a trade-off between trust centralization and overhead.
Additionally, we limit the frequent time-consuming access to the blockchain-based storage system by explicitly separating the real-time dynamic (or behavioral) data and the less dynamic (or static) data~\cite{suhail2020trustworthy} due to time-sensitive tasks in CPSs. This strategy is adopted due to the fact that the real-time data evolves frequently with each lifecycle phase and usually amasses in intervals ranging from minutes to milliseconds, such as deployed sensors and actuators data providing asset's actual state, while static data changes infrequently with time along the lifecycle of the real-world counterpart, such as provenance data, system historical data, device configuration settings, access levels, and policies. Thus, in addition to application- or user-specific requirements, critical data and sources that can facilitate the track and trace solutions for data-driven CPS must be considered.
The selection of apropos blockchain further depends on scalable solutions that become critical due to the growing number of actors and activities in the CPSs. To enforce scalable solutions, on-chain solutions based on the data structure of blockchain, i.e., DAG-structured blockchains, are better options than chain-structured blockchains. In terms of implementation, most of the DAG-structured blockchains may not be mature enough for mass adoption yet. 
Therefore, chain-structured blockchains, for instance, Hyperledger, with the cost of off-chain solutions to address scalability can be adopted. However, most of the chain-structured blockchains are not quantum-safe.  

In the following, we discuss safety and security (S\&S) rules, one of the significant and critical components for securing CPSs through the blockchain-based DT framework.
S\&S rules are generated based on threshold data (upper and lower bounds), consistency checks (pre-defined performance parameters), and fine- and coarse-grained constraints (data accessibility and auditability based on ownership, roles, and access levels).
S\&S rules can be defined at device level, such as device configuration details, predefined performance parameters, network level, such as topology, communication path, or process-level, such as data aggregation, relationship between entities. S\&S rules support monitoring the actions and events of the participating entities inside the virtual environment, thereby supplementing safety instrumented systems (SIS).
Safety rules, for instance, conditional limits for device data (e.g., minimum and maximum temperature), trends/patterns (e.g., excessive vibration or heat), consistency checks (speed-variable), etc. and security rules, for instance, data integrity, access control (authentication and authorization), or correlation of different DTs associated with different physical sub-components can be stated and integrated. Such IT/OT-system specifications are represented by means of standardized data formats, such as Automation Markup Language (AML), that consider the syntactical and semantic levels to describe the data objects~\cite{eckhart2018towards}.


Malicious or accidental, S\&S rules must be integrated to detect misconfigurations and to mitigate the outcomes of malfunctioning components and malicious activities in a CPS. Inability to impose such rules leads to devastating results on the system in the form of cyberattacks or APTs.
Furthermore, introducing S\&S rules at the design phase can help to lower security and incident response costs, thereby making later lifecycle phases less prone to errors and incidents~\cite{dietz2020unleashing}.
The S\&S rules can be deployed through smart contracts (executable code running on the blockchain) to invoke the appropriate defense mechanisms automatically~\cite{suhail2021blockchainbased}. Additionally, to track an accountable entity generating or updating rules for a given asset are also recorded. 
Depending on the cyber situation, S\&S rules can be updated either during or after the operation or process. For instance, in any abnormal event during the ongoing process, the system needs to respond effectively and update the corresponding S\&S rules to avoid long-term loss. 
To strengthen the rationale for integrating blockchain with DTs, the S\&S rules must be stored and retrieved from the ledger to ensure their reliability. For instance, a provenance-aware blockchain-based system can track and trace the accountable entity by adding or updating the S\&S rules.  


\subsection{Combining Digital Twins with Threat Intelligence}

{\textbf{Challenge \#4:}} {\it Why we need TI in the presence of DTs for predictive maintenance and situational awareness?} 

While DT security-operation modes offer a variety of advantages, they may pose limitations. 
For the replication mode, the time-dependent {\it record and replay} of states may not create future scenarios. The synchronization issues between the physical object and its different replicas are essential whereas the state replication accuracy depends on the trade-off between budget and fidelity~\cite{bitton2018deriving}. Additionally, the input knowledge (events) is required in advance to produce the same stimuli.
Similarly, for simulation mode, since the current physical state of the system is not known, hence it has to rely on user-specified settings and parameters. Considering these shortcomings, we envision how TI provides auxiliary support for predictive maintenance and situational awareness.  


Creating future scenarios, particularly in the absence of direct observation data, is challenging for DTs~\cite{rasheed2020digital}. To predict anomaly detection and perform risk assessment, the predictive capability of ML-based solutions available through the TI module can play a key role. Thus, TI can provide additional support for DTs to secure CPSs. Furthermore, the {\it quality} of the data matters at par with {\it quantity} for ensuring precise predictions and decision-making. Such requirement matters more in the presence of a complex and ever-changing threat landscape and high volume, velocity, and variety of big data.
Therefore, we propose a TI module that can learn useful patterns from the collected big data to prevent/detect anomalies and to predict the system behavior in course of disruptions or thwart the attack cycle.
Furthermore, the TI module can be integrated into the organizational security management framework, such as SIEM~\cite{dietz2020integrating} (as shown in Fig.~\ref{fig:abstract}), to deduce and to check the adherence of security rules. The TI data set can also provide valuable clues (such as unknown IP address or unusual network traffic) to carry out a system-wide search for bad actors during threat hunting.


Data-driven TI exploits data from DTs security operation modes to achieve predictive powers and situational awareness.
In the following, we explore how DT security operation modes operate in combination with TI and under which settings such modes are desirable?
The replication mode along with TI can perform Prognostics and Health Management (PHM) where DT of the equipment supported by data is used to detect failures or repair needs. By using ML algorithms such as Long Short-Term Memory (LSTM) and Support Vector Machine (SVM), the condition of the given asset is predicted based on performance constraints or tool wear data~\cite{groshev2021toward, rasheed2020digital}. If the asset state data might lead to failure situations, it schedules the maintenance of the physical asset. 
TI can access data from data sync. The reason for accessing data from this module is that it mirrors the sync states of both physical and virtual objects, hence can analyze the behavior of both spaces.
While TI predicts existing and potential threats, data sync can also retrieve data from TI to figure out the reason for data inconsistencies.  

The simulation mode along with TI can serve as a training and testing platform to improve the security of the infrastructure. 
Through DTs, attack scenarios can be simulated to analyze the system behavior under attack. Collecting data during such events can help to derive patterns to be formulated, tested, and transmitted to real-world systems. Moreover, DTs can aid security professionals by providing red-blue team exercises for cybersecurity training opportunities~\cite{becue2018cyberfactory} (i.e., the offensive red team launch attacks to uncover system weaknesses while the defensive blue team implements adequate solutions). Together with TI, it can predict the possibility of attacks or system malfunctioning (i.e., risk assessment) to carry out what-if and cost-benefit analysis. To do so, ML algorithms such as K-means and autoencoders are ideal solutions for traffic inspection~\cite{groshev2021toward}. Based on the inputted parameters that are mapped to the training data, TI learns about the presence of an attack or abnormal situation, and hence generates S\&S rules. These rules are then verified on the testing data.
Thus, DT replication modes together with TI can gain insights into the root cause that leads the system to unexpected behavior.


In the case of a known threat or attack, alerts are triggered to take further actions, such as executing S\&S rules to activate the corresponding mitigation strategies. 
In the case of an advanced stealthy threat or attack, TI may not be able to generate an alarm instantly. Such prediction errors are caused due to model errors coupled with the bias in the data. Under such circumstances, depending on attack intensity, either the device or network log data can be analyzed by using threat hunting to identify the root cause of attacks or switch off the affected device or service. 
 
\begin{figure}[!t]
\centerline{\includegraphics[width=3.0in]{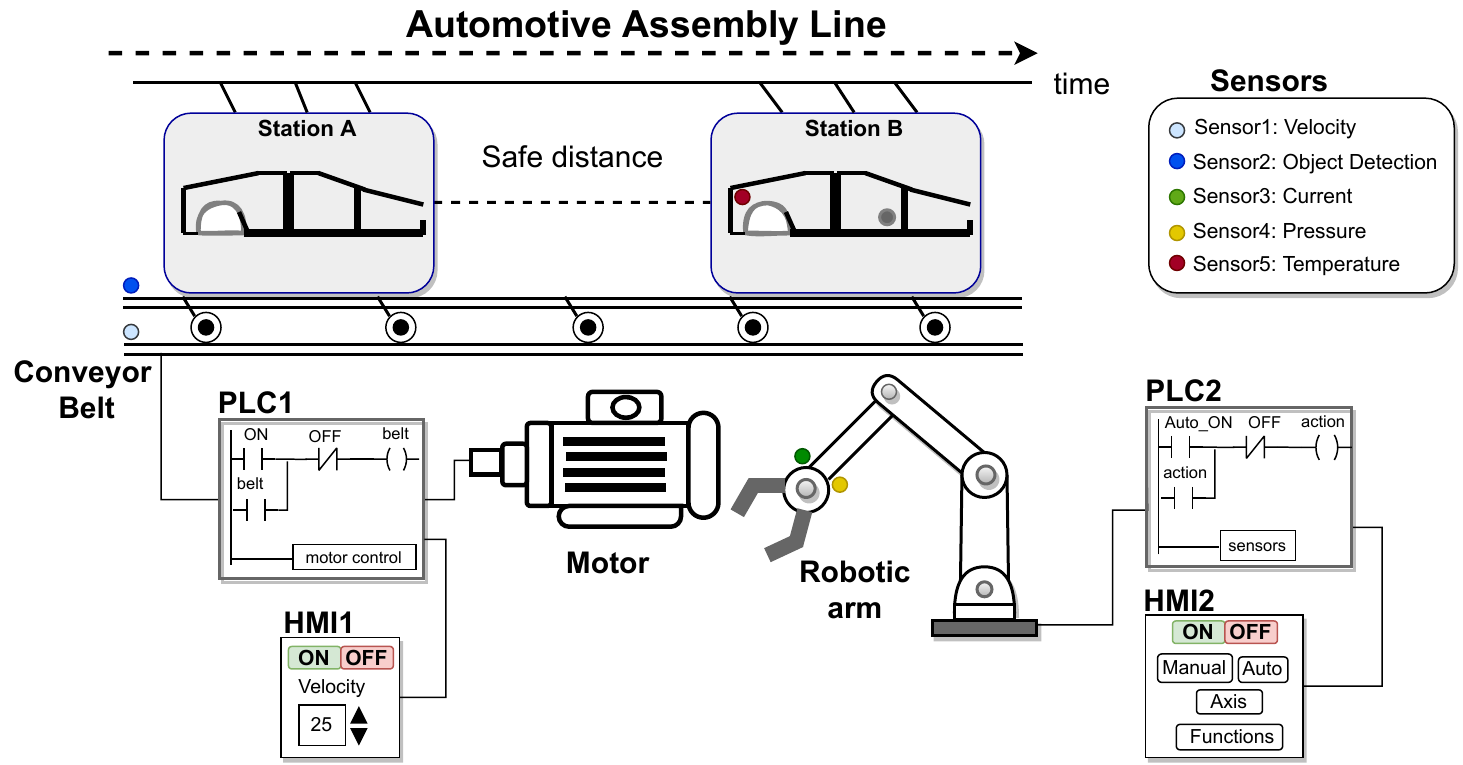}}
\caption{Automotive assembly line: A scenario specification.}
\label{fig:scenario2}
\end{figure}

\section{A CPS Use Case: From Design to Dismissal} \label{sec:use_case}
This section discusses a CPS use case of an assembly line in the automotive industry where we map an exemplary physical process (Fig.~\ref{fig:scenario2}) to our proposed framework (Fig.~\ref{fig:framework}). 
The assembly line has multiple stations equipped with machinery for performing dedicated tasks. The motor-driven conveyor system can move the objects (chassis) from Station A (chassis loading point) to Station B (chassis welding point) for performing a physical task (such as spot welding and parts assemblage). Initially, the motor is off. Firstly, the chassis are required to be loaded on the belt at Station A and their presence is being detected by a proximity sensor (Sensor1: Object Detection). To do so, the velocity (Sensor2: Velocity) of the conveyor system must be monitored against a certain threshold based on the following conditions: (i) detect and load only a specific number of chassis on the belt, and (ii) maintain a safe distance to avoid collision between two adjacent chassis on the assembly line. Based on a predefined task duration, the chassis at Station A can be moved to Station B, while more chassis can be loaded at Station A. Secondly, the welding operation on the chassis is performed by a robotic arm at Station B. To do so, the welding gun applies appropriate current and pressure at the welding spot measured through Sensor3: Current and Sensor4: Pressure. Another sensor (Sensor5: Temperature) measures the temperature during the welding process and is bounded by a threshold to avoid material deterioration. 
To monitor tool wear data, the robotic arm is equipped with sensors (such as vibration, force cutting, and acoustic emission). During manufacturing processes, machine unavailability (due to equipment deterioration or machine malfunctioning) and uncertain disturbances (due to urgent job arrival or job tardiness) usually occur, leading to performance and production disruption. Therefore, we keep monitoring tool wear data of the robotic arm. Recording such data ensures continuous assembly line operation while triggering a suitable time for machine maintenance or timely rescheduling for dynamic job-shop scheduling. We calculate tool wear data based on the number of objects (chassis) being welded by the robotic arm. 
To command and control physical processes (for instance, turning the motor and robotic arm on/off, setting velocity of the motor, or setting current and pressure of the robotic arm, etc.) through the belt and robotic arm PLC1, HMI1, PLC2, HMI2 are used.

Before starting the production process, the necessary details mentioned in Section~\ref{DDI} can be performed as follows: 
(i) The participating entities (sensors, actuators, robotic arm, PLCs, HMIs, etc.) registered at the blockchain (Fig.~\ref{fig:framework} {\it step 1}); (ii) The input sensory data (Fig.~\ref{fig:framework} {\it step 2a}) is mapped to the ICMs (Fig.~\ref{fig:framework} {\it step 2b}); (iii) The auxiliary data (Fig.~\ref{fig:framework} {\it step 3a}) such as, supply chain data including consignment information, material stock, production quantity, etc. is fused with the unified sensory data (Fig.~\ref{fig:framework} {\it step 3c}). The twined data from other physical processes, for instance, assemblage of components, may also be integrated (Fig.~\ref{fig:framework} {\it step 3b}). Furthermore, for tracking data generating source, the provenance data is also stored in blockchain (Fig.~\ref{fig:framework} {\it step 4}). 

To monitor and analyze the production processes, depending on the application-specific requirements, the DTs security mode operations, i.e., simulation and/or replication can be used. 
Through the simulation mode, we may achieve optimal operating conditions for the following subprocesses (i) how to maintain a safe distance between adjacent chassis on the conveyor belt?, (ii) how to maintain a temperature bounded by a threshold to avoid material deterioration during the welding process by robotic arm?, and (iii) how to deal with incident handling during assemblage of vehicle parts performed by a robotic arm? All such tasks must be monitored and configured first based on simulation setting artifacts to deal with incident handling (Fig.~\ref{fig:framework} {\it step 8}). Along with TI, the incident data (including log data, provenance data, etc.) can evaluate the risk assessment in terms of detection reliability and severity to enable incident investigation (Fig.~\ref{fig:framework} {\it step 7b}). Based on the incident data, S\&S rules are constructed or improved and are updated on blockchain (Fig.~\ref{fig:framework} {\it step 9a} and {\it step 9b}). Thus, enforced through S\&S rules, such proactive strategies can help detect safety- or security-related attacks instantly during the actual manufacturing of vehicles and hence can radically reduce maintenance overheads to alleviate product defects and support secure and trusted manufacturing. 


Through replication mode, we may achieve mirrored operating conditions for the following subprocesses (i) how to monitor the velocity of the motor-driven conveyor system against a certain threshold (defined in S\&S rules) to avoid attacks (such as MITM) targeting the manipulation of the conveyor speed?, (ii) based on tool wear data how to estimate the asset capacity for the next production process?, and (iii) based on machine performance data how to to predict fault diagnosis to minimize the makespan and production cost? To do so, the data sync method makes use of the input knowledge from (i) the live physical asset current state or events recorded through sensors (Fig.~\ref{fig:framework} {\it step 6a}) and (ii) ICMs from blockchain (Fig.~\ref{fig:framework} {\it step 5}) to reproduce the same stimuli in DTs (Fig.~\ref{fig:framework} {\it step 6b}). Furthermore, data sync keeps replicated state information (Fig.~\ref{fig:framework} {\it step 6c}) that can be access by TI (Fig.~\ref{fig:framework} {\it step 7a}) to continuously monitor the erratic behavior of ongoing operations based on the relationships between the dynamic variables and historical variables for any rule violations. 

The process of collecting and analyzing data continues even after the vehicle production process. To optimize operational control, fault diagnostics, and prognostic health management, the sensors attached to the connected vehicle (such as camera, GPS, infrared detectors, etc.) collect data during or after the drive. The accumulated data is then fed to the DTs, which self-adapts to the asset and extracts new knowledge for model calibration for the next production processes. Furthermore, such data can be used by TI to study the adversarial space and to update S\&S rules for detecting advanced attacks.

The dismissal of the object is followed by the decommissioning of the DT because of obsolescence or other reasons. At this stage, the product lifecycle data are backed up and made available to other objects or domain experts to optimize the production of future devices~\cite{suhail2021blockchainbased}.


\section{Concluding Remarks and Future Directions} \label{conclusion}
This article focuses on securing CPSs by integrating TI and blockchain for intelligent and trusted DTs. Implementing intelligence-driven architectures enables greater visibility of cyber threats and minimizes the threat landscape; however, it necessitates methodological and theoretical solutions to address longstanding challenges. For instance, 
\begin{itemize}
\item Being the virtual replicas of their physical counterparts, DTs share commonalities in implementation and operational behavior. Due to this fact, DTs may act as a source of data breach, leading to the abuse case of DTs~\cite{Eckhart2019} where attackers may launch covert attacks on the physical system by exploiting the valuable system knowledge available through DTs. For example, attackers may sabotage the system by manipulating DTs data such that the physical counterpart may not indicate data inconsistencies. Attacks instigated through DTs may have similar repercussions as that of directly attacking real field devices. This problem may raise the question: how the security of DTs can be accessed?
One possible solution to assess the security of DTs is to launch attacks on DTs from the cyber range (i.e., virtual environments that provide hands-on cyber skills and security posture testing). Some of the pointers for this future research direction include gamification strategies that not only serves for training and testing purposes~\cite{vielberth2021digital, becue2018cyberfactory} but can also evaluate the security of DTs and eventually physical asset. Furthermore, blockchain-based DTs can define data accessibility and auditability based on ownership, role, and access levels.  


\item DTs must exhibit sufficient fidelity to protect against failures of the real CPS. However, uncertain scenarios may occur due to the dynamism and complexity of underlying (sub) systems, for instance, machinery malfunctioning, hidden vulnerabilities, erroneous data, knowledge gaps, etc. This problem may raise the question: how accurately DTs are required to follow the states of their physical counterparts? Determining the desired fidelity of DTs for realizing the underlying use case requires further investigation with the focus on cost-efficient development of DTs~\cite{bitton2018deriving}. Additionally, relying on the proposed framework (Fig.~\ref{fig:framework}) can help to obtain the process knowledge from DTs as well as ML methods to learn security- and privacy-related aspects based on sensor data. 

\item On the one hand, due to the black-box nature of AI models, the decisions made by the resulting systems suffer from a loss of transparency and comprehensibility, hence leading to a negative impact on the system's trustworthiness. However, on the other hand, complete transparency into the inner working of AI models may expose them to adversarial attacks by allowing them to make inferences from live cyber data or executing model poisoning into the training workflows. Attackers may begin with AI reconnaissance (i.e., observe the model's behavior and learn the decision logic or knowledge bases to morph the malicious payloads) and, without being detected, can feed their adversarial data to launch data exfiltration attacks.
In this regard, blockchain can support transparency for AI operations by assuring the traceability and solidity of data and algorithms, thereby establishing more understanding and confidence in the decisions made by the underlying systems~\cite{dinh2018ai}.
\end{itemize}

The proposed framework highlights various security-enhancing use cases of DTs in conjunction with TI, albeit it has the following limitations: (i) how the proposed framework can still work if a well-versed adversary steers the CPS into an insecure state by manipulating the benign behaviour of DTs or exploiting the cyclic state update (from the physical process to DTs or vice versa) during replication mode, (ii) how to improve the resilience of underlying physical processes during an attack, and (iii) most importantly, investigating the practicality of the concept in a real-world setting. 



\end{document}